\documentstyle[11pt,epsfig]{article}
\def\be{\begin{eqnarray}}
\def\ee{\end{eqnarray}}

\def\J#1#2#3#4{ {#1} {\bf #2}, {#3} (#4) }
\def\PRL{Phys. Rev. Lett.}

\def\PLB{Phys. Lett. B}

\def\NPA{Nucl. Phys. A}
\def\NPB{Nucl. Phys. B}

\def\PRC{Phys. Rev. C}
\def\PRPTS{Physics Reports}
\def\bea{\be}
\def\eea{\ee}

\def\roughly#1{\mathrel{\raise.3ex\hbox{$#1$\kern-.75em%
\lower1ex\hbox{$\sim$}}}}
\def\lsim{\roughly<}
\def\gsim{\roughly>}

\def\la{\langle}\def\ra{\rangle}
\def\PRL{Phys. Rev. Lett.}

\def\PLB{Phys. Lett. B}

\def\NPA{Nucl. Phys. A}
\def\NPB{Nucl. Phys. B}

\def\PRC{Phys. Rev. C}

\begin{document}

\renewcommand{\thefootnote}{\fnsymbol{footnote}}
\setcounter{footnote}{0}

\vskip 0.4cm \hfill {\bf KIAS-P99091}

\hfill {\today}
\vskip 1cm

\begin{center}
{\LARGE\bf Selected Topics in Astro-Hadron Physics:} \vskip 0.2cm

{\LARGE\bf Going from a Proton to Nuclei to Neutron
Stars}\footnote{Note prepared for the 4th Bulletin of Asia Pacific
Center for Theoretical Physics, Seoul, Korea}

\date{\today}

\vskip 1.5cm

{\large
Mannque Rho
}

\end{center}

\vskip 0.5cm

\begin{center}

{\it Service de Physique Th\'eorique, CE Saclay, 91191
Gif-sur-Yvette, France}

{E-mail: rho@spht.saclay.cea.fr}

{and}

{\it School of Physics, Korea Institute for Advanced Study, Seoul
130-012, Korea}

\end{center}

\vskip 0.5cm

\begin{abstract}

The effort currently in vogue in some small circle of physicists
to go from a proton to nuclei to compact stars involves various
aspects of particle and nuclear physics that require input from
laboratory experiments, transcending narrow specialization in
diverse sub-fields. Several topics on this matter are discussed in
this note. The notion of Cheshire Cat Principle is introduced for
nucleons, nuclei and dense hadronic matter and is confronted with
experimental data on the proton, two-nucleon systems and heavy-ion
experiments, with a leaping extrapolation to the structure of
neutron stars. The matter discussed here illustrates that a close
contact with experiments, indispensable in the present case, is
essential for significant progress in {\it any} field of physics.
I suggest that this is a field that has a tremendous potential for
breakthrough in the Asia Pacific countries, particularly in Korea,
where some of the seminal works in this area have been done by
young theorists, both pre- and post-graduate.

\end{abstract}

\newpage

\renewcommand{\thefootnote}{\#\arabic{footnote}}
\setcounter{footnote}{0}

\section{Introduction}\label{intro}
\indent\indent A relatively new branch of physics combining
hadronic physics and astrophysics is called ``astro-hadron
physics." I would like to describe some recent efforts to
understand such extremely compact astrophysical objects as neutron
stars starting from data obtained in laboratories, in particular
from heavy-ion experiments that have been performed in such
laboratories as GSI of Germany, CERN (SPS) of Switzerland and
elsewhere. The central thesis that I will develop is that there
are what may look like ``alternative" descriptions for the same
process in different languages ranging from bare hadrons to quasiparticle
hadrons to quasiparticle quarks characterized by what I would call
``Cheshire Cat mechanism." This ``duality" nature seems to be operative
from elementary hadrons to superdense matter perhaps existing in
compact stars, suggesting a unity of particle, nuclear and astro physics.

As in all works at an embryonic stage, there are false
starts, pitfalls, wrong tracks and controversies. What I shall
describe therefore are not necessarily well-established facts.
Much work will be needed to validate or invalidate some or all of
them. Even so, what we have at this stage is so exciting that it
deserves much more attention than presently paid, particularly
from the Asia Pacific physics community where some of the early
significant works have been done by young theorists working on PhD
theses~\footnote{Some of the earlier developments were already 
discussed in the 1997
APCTP-sponsored workshop on astro-hadron physics held in 
Seoul~\cite{97astro}. As a follow-up and for more updated developments,
a workshop is planned for the year 2000 at Korea 
Institute for Advanced Study (KIAS) in Seoul devoted to 
such phenomena as supernovae, 
neutron stars, black holes, gamma-ray bursts etc.}.

I will start the discussion with the proton, the constituent of
the nucleus, go to the simplest nucleus, i.e., two-nucleon system
including the deuteron, then to nuclear matter and finally to
dense hadronic matter relevant to the interior of neutron stars.
The topics are selected to highlight work done in Korea by thesis
students or young theorists. Clearly part of the connections are
incomplete or faulty, needing more solid structure but the logic
appears sound and worth pursuing for possible breakthrough.
\section{``Proton Spin Problem"}
\indent\indent I start with the nucleon, specifically the proton
which is the lowest baryonic state. In the fundamental theory of
strong interactions, QCD, this state is a bound state of three
quarks in the color-singlet state confined in as yet poorly
understood way within a region wherein the gluons play a crucial
role. To describe this from the QCD Lagrangian using its
microscopic variables is at present practically impossible but the
most tantalizing fact is that it can be given various different
(albeit approximate, yet qualitatively correct) descriptions
indicating some sort of ``dualities" are in action. One apt way of
seeing this is to use ``Cheshire Cat principle" (for a general
introduction, see \cite{NRZ}).
\subsection{Cheshire Cat Principle}
\indent\indent
 Consider three colored quarks $uud$ of the quantum numbers of
 a proton confined in a spherical
``bag" of radius $R$ surrounded by a cloud of Goldstone pions, the
latter being indispensable to the system in order to be consistent
with the spontaneously broken chiral symmetry. With a suitable set
of boundary conditions consistent with the symmetries of QCD that
mediate the communication between the inside microscopic QCD
variables and the outside macroscopic hadronic variables, nearly
all low-energy observables of the proton can be understood
independently of the size of the confining bag~\cite{NRZ,toki}.
Indeed for a large bag with $R\sim 1$ fm, the quark-gluon degrees
of freedom dominate while for a small bag with $R\lsim 1/2$ fm, it
is the Goldstone degrees of freedom that govern the dynamics of
the system. This feature is known as the Cheshire Cat
phenomenon~\cite{cheshire}. When the ``bag" is shrunk to a point
$R\rightarrow 0$, which is allowed by Cheshire Cat principle, one
gets the celebrated skyrmion~\cite{skyrme,witten,NRZ}, a
description of the baryon that is the more accurate the larger the
number of colors $N_c$. It seems that in nature, $N_c=3$ is
already quite large, so the skyrmion picture is qualitatively
correct.
\subsection{Flavor-singlet axial charge}
\indent\indent
 One conspicuous exception to the success of the
Cheshire Cat phenomenon has, up to date, been the flavor singlet
axial charge (which will be referred in short to as FSAC) of the
proton which is often associated with the so-called ``proton-spin
problem." It turns out that there is no obstacle to the Cheshire
Cat manifesting in this quantity, that is, there is no
``proton-spin crisis." The resolution lies in subtle quantum
anomalies which are now understood. We do not yet have a complete
answer to the issue which is the PhD thesis subject of Hee-Jung
Lee at Seoul National University but I will briefly describe how
this Cheshire Cat property can be recovered in the FSAC when
chiral symmetry and chiral anomaly are judiciously taken into
account~\cite{heejung}. The interplay between the boundary
conditions and Casimir effects is found to play a crucial role.

\begin{figure}[ht]
\centerline{\epsfig{file=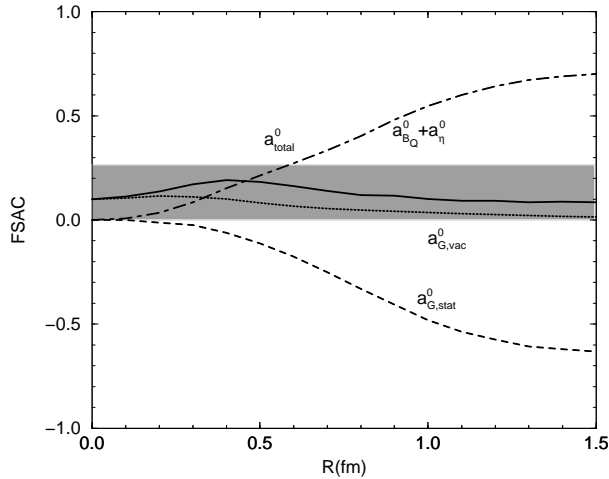, width=9cm}} \caption{The flavor
singlet axial charge of the proton as a function of the bag radius
compared with the experiment; it consists of three contributions:
(a) {\it matter field contribution}: quark plus $\eta$ ($a^0_{B_Q}
+ a^0_\eta$), (b) {\it gauge field contribution}: the static
gluons coupled to the quark source ($a^0_{G,stat}$), (c) {\it
Casimir contribution}: the gluon vacuum fluctuation
($a^0_{G,vac}$), and (d) the sum total ($a^0_{total})$. The shaded
area represents the range admitted by experiments.}\label{fsac}
\end{figure}

Since the flavor-singlet axial current is not conserved because of
the anomaly, the color cannot be confined inside the bag unless a
suitable boundary condition is put at the surface that cancels the
outflow of the color as discovered by H.B. Nielsen et
al~\cite{NRWZ}. The boundary term that does this is proportional
to the Chern-Simons current on the surface, i.e., the Chern-Simons
flux (which is invariant under neither small nor large gauge
transformation). This influences nontrivially the FSAC of the
proton. In a nut-shell, what happens is that the FSAC contributed
by the matter fields (quarks inside the bag and $\eta^\prime$
outside the bag) and the FSAC contributed by the gauge field
(gluons inside the bag) more or less (or possibly exactly if
treated rigorously) cancel, leaving behind only the small
contribution from the (gauge field) vacuum fluctuation which is
effectively a Casimir effect caused by the boundary with its
color-anomalous boundary condition. The cancellation and the
remnant small FSAC are shown in Fig. \ref{fsac}.

The small FSAC of the proton that is left over -- which is
independent of the size $R$ -- provides another evidence that the
proton could be equivalently understood {\it both} in terms of
quarks/gluons and in terms of macroscopic hadronic variables. When
$R$ is taken to be big, it is the QCD variables that figure
significantly, e.g., the MIT bag. When the size $R$ is shrunk to a
point, the proton is a skyrmion. Thus we have the equivalence of
the skyrmionic proton and the quark-gluonic (QCD) proton, that is,
the Cheshire Cat. Since there is no way that one can exactly
bosonize four-dimensional QCD, the equivalence can be only
approximate. Thus we simply have an approximate equivalence which
can be made more precise by doing more work.

In what follows, I will develop the thesis that it is this feature
observed in the proton that underlies many-body dynamics going all
the way to neutron star matter.

\section{Nuclei in Effective Field Theory}
\indent\indent
 Let me now turn to nuclei. For this, consider two
nucleon interactions at very low energy. The two-nucleon system is
the nuclear system that is the simplest -- and the only -- nuclear
system that can be treated accurately and systematically. At
present it is only at the probed energy-momentum much less than
the pion mass $m_\pi\sim 140$ MeV that is amenable to an accurate
computation . At very low energy, according to Weinberg's
``theorem," the content of QCD can be phrased in terms of the
nucleons and pions treated as effective fields. In fact if we are
probing a scale much less than the pion mass, we can even ignore
the pions and work with the nucleons only. The corresponding
framework is an effective field theory (EFT). What this means is
that we can shrink -- following the notion of the Cheshire Cat --
the bag of the nucleon and work in the skyrmion limit, namely,
zero-radius bag. Thus a nucleus will be a collection of point-like
objects interacting strongly subject to the constraint of QCD
symmetries.

There has recently been an intense activity on this EFT for
two-nucleon systems~\cite{weinberg,pmr,vankolck,KSW,PKMR}. Some of
the developments are summarized in a series of workshops devoted
to the issue~\cite{EFTmeetings}. Up to date, there are two
successful approaches to EFT in nuclear physics. One is the
original Weinberg approach~\cite{weinberg} where a systematic
power counting is made only to the ``irreducible graphs,'' for
which chiral perturbation theory (with pions figuring prominently)
becomes applicable in organizing the expansion of the series and
the reducible graphs are summed to all orders with the irreducible
graphs entering as vertices. This scheme used in
Ref.\cite{weinberg,pmr,vankolck,PKMR} -- which in spirit is close
to the original Wilsonian EFT but incurs possible errors in the
power counting -- involves a chiral-symmetry scale $\Lambda\gsim
m_\pi$ (where $m_\pi\sim 140$ MeV is the pion mass) as the
counting is applied only to the irreducible terms. I will call
this the $\Lambda$ counting. The other approach~\cite{KSW}
motivated to account more explicitly for the anomalously large
s-wave scattering lengths in the two-nucleon scattering allows one
to do a systematic counting for the S-matrix as a whole. This
approach renders a more systematic accounting of the powers of
$Q/\Lambda$ where $Q=\sqrt{MB}$ (where $B$ is the deuteron binding
energy and $M$ is the proton mass), $p$ (probe momentum) as well
as $m_\pi$ (pion mass) but at the expense of certain predictivity.
This is referred to as $Q$-counting scheme.

In essence, both approaches, though somewhat different in
strategy, are equally consistent with the tenet of EFT and more or
less equivalent in their predictive power. In what follows, I will
simply focus on the $\Lambda$-counting approach which works
stunningly well for two-body systems. Actually in the processes
that I will consider, the $\Lambda$ counting is found to be more
readily adaptable to
--- and predictive in treating -- nuclear physics problems. One
non-trivial advantage of the $\Lambda$ scheme is that it allows
one to calculate precisely defined corrections to what can be
obtained from so-called realistic potential models (PM in short)
that have been developed by nuclear theorists since a long time,
thus giving the realistic potential models (PM) a first-principle
justification. It allows us to study processes involving not only
few-body but also many-body systems. For instance, it is possible
to calculate the ``hep" process in the Sun $p+^3{\rm
He}\rightarrow ^4{\rm He}+e^++\nu_e$ (which is currently an
exciting issue after the recent Surperkamiokande neutrino data)
with an accuracy that can be controlled systematically. Such a
calculation in the $Q$-counting scheme is most likely to be a hard
task.

In a series of recent papers, Tae-Sun Park (currently a 
post-doc at TRIUMF in Vancouver) and his
collaborators~\cite{PKMR} have shown in a finite-cutoff
regularization~\footnote{The cutoff is {\it not} to be sent to
infinity in EFT contrary to renormalizable field
theories~\cite{lepage}.} that the EFT results of the leading order
terms in {\it all} two-body observables at low energy $E\ll m_\pi$
are precisely reproduced by the potential model results. This is
the case not only for scattering amplitudes but also all
electroweak response functions. What EFT can do that the potential
models (PM) cannot is that the corrections to the leading order
results given accurately by the PM are calculable systematically.
This is the {\it power} of EFT as a {\it theory}. For low-energy
processes, this privileged role of the PM in EFT can be understood
by the fact that the tail of the wave functions is a physical
quantity and the realistic potential models which are fit to
experiments have the {\it correct} asymptotic properties in the
wave function~\cite{pcohen}.

Considered to order $Q^n$ where $n$ is the order in the $\Lambda$
counting (which I will consider relative to the leading order term
in the expansion of the irreducible graphs), the s-wave scattering
amplitudes are accurately postdicted 
by T.-S. Park et al~\cite{PKMR} and further improved 
by Chang-Ho Hyun (a graduate student of Seoul National
University) et al~\cite{hmp} up to $p\leq
m_\pi$ for $n=2$ and a cutoff appropriate to the number of pions
exchanged (one or two) in the irreducible graphs. All deuteron
properties are also well understood within the same
scheme~\cite{PKMR}.  As an important spin-off, the scheme allowed
the calculations to order $Q^2$ and $Q^3$ of the proton fusion
process in the Sun~\cite{pp} \bea p+p\rightarrow d+e^+
+\nu_e\label{pp} \eea and of the threshold np
capture~\cite{pmr,pkmr99} with polarized projectile and target
nucleons
\bea
 \vec{n}+\vec{p}\rightarrow d+\gamma.\label{np}
 \eea
The process (\ref{pp}) crucial for the solar neutrino problem is
given in the scheme to an accuracy of $1\sim 3$ percents (the
uncertainty here is due to the exchange current that appears at
order $Q^3$). The unpolarized cross section for (\ref{np}) has
been computed to the accuracy of 1 percent in complete agreement
with the experiment. More significantly, the polarization
observables $P$ (circular polarization) and $\eta$ (anisotropy)
have been predicted {\it parameter-free} in Ref.\cite{pkmr99,CRS}.
Since there are no experimental data available yet, this is a
genuine prediction involving matrix elements that are suppressed
relative to the allowed term by $\sim$ 3 orders of magnitude.
These quantities are currently being measured in several
laboratories and will soon be available. The outcome will be an
exciting check of the prediction of the theory, perhaps the first
of the kind in nuclear physics.

In all these postdictions and predictions, there is very little
$\Lambda$ dependence as required by the tenet of EFT, assuring
that the scheme is fully consistent.

 One can go up in the momentum range by doing higher
order calculations. Phillips and Cohen~\cite{pcohen} discuss how
the two-body EM form factors can be described in the $\Lambda$
scheme. Pushing somewhat the validity of the scheme, one can
calculate even the process
\bea
 e+d\rightarrow e+n+p
 \eea
involving large momentum transfers $q\gsim 1$ GeV. In fact this
process measured in 1980's at ALS of Saclay and elsewhere is
considered to be the {\it unambiguous} confirmation of
meson-exchange currents in nuclei (see ~\cite{rhoarnps,frois}).

In sum, the dilute nuclear systems such as the two-nucleon bound
and scattering states can be quite accurately described in terms of
certain effective degrees of freedom connected to QCD via a
Cheshire Cat mechanism. I shall now jump directly to the
infinite-body problem, namely, nuclear matter, although the
extension of EFT to three- or more-body problems is not fully
worked out yet.
\section{Dense Hadronic Matter: BR Scaling}\label{brscaling}
\indent\indent
 For more than a few nucleons and many-nucleon
systems like nuclear matter and denser matter, the EFT described
above cannot be straightforwardly applied. In fact a systematic
approach of the type does not yet exist. The reason is rather
simple to understand. First of all when several nucleons are
involved the relevant kinematics is not always one to which the
low-energy/momentum expansion with a manageable number of terms is
applicable and further a systematic expansion would involve many
terms whose coefficients are not fully determined from either
theory or experiments. This means that strictly speaking, no
parameter-free calculation that does not invoke some ad hoc
assumptions can be done. Thus some clever intuition is needed to
overcome this technical difficulty. One economic way to short-cut
the formidable-looking obstacles is the BR scaling introduced by
Brown and Rho~\cite{BRscaling}. Clearly this cannot be the only
way but it has not yet met with contradictions while having a
variety of success.

The basic idea is that nuclear matter at its equilibrium density
represents the ground state of the matter and that at that point,
the nucleons and mesons are quasiparticles as they are at zero
density. We know from Migdal's work that nucleons in nuclear matter
are quasiparticles in the sense of Landau Fermi liquid
theory~\cite{migdal}. In modern language, this means that the
nucleon mass and quasiparticle interactions are fixed-point
quantities. We go one step further and assume that mesons are also
quasiparticles at the equilibrium point of nuclear matter. This may
sound absurd as one knows that mesons interact strongly with many
inelastic channels open, so the notion of quasiparticles with
well-defined {\it effective} mass for them does not sound right.
But then before the advent of shell model one used to say the same
thing about the nucleons and yet nucleons in nuclear matter are
bona-fide quasiparticles. Skeptics will then argue that the
quasiparticle notion works for the nucleons because of Pauli
exclusion principle but for bosons, there is no such thing. This is
a valid objection to which no one can at present offer satisfactory
answers. So assuming that bosons in medium can be treated in the
tree approximation with a point-like structure is an assumption yet
to be tested.  I shall simply proceed to use the notion in a
variety of processes until it meets contradictions. So far there is
no evidence that this notion conflicts with nature.

The next step then is to construct an effective Lagrangian theory
which preserves the known symmetries of QCD. In principle, this
Lagrangian should be usable for a systematic calculation of the
type described above for two-nucleon interactions and indeed for
particles near ``on-shell" in medium, such higher-order
calculations must be done to describe the needed amplitudes. I
shall focus, however, on processes which are off-shell and hence
can be treated in the tree order.

What are then the relevant degrees of freedom? We assume that
Goldstone theorem is applicable in medium, that is, there are
zero-mass Goldstone (in the chiral limit) or light-mass
pseudo-Goldstone (in nature) particles $\pi^a$, massive nucleons
$N$ appearing as matter fields, vector bosons $V_\mu$ and possibly
scalars $\sigma$ etc. We assume that such classic low-energy
``theorems" as Goldberger-Treiman, Adler-Weisberger,
Gell-Mann-Oakes-Renner, Kawarabayashi-Suzuki-Ryazuddin-Fyyazuddin
... relations hold in the medium but with masses given by
\bea
 \frac{m_N^*}{m_N}\approx \frac{m_V^*}{m_V} \approx
\frac{m_\sigma^*}{m_\sigma}\approx\cdots\approx
\frac{f_\pi^*}{f_\pi}\approx
(\frac{\la\bar{q}q\ra^*}{\la\bar{q}q\ra})^n.\label{BR}
 \eea
  Here
the star denotes in-medium quantity and $\la\bar{q}q\ra$ stands
for the quark condensate in the vacuum (the starred quantity being
the same in the in-medium ``vacuum.") The index $n$ depends on
models. Empirically it is close to $1/2$. The quark condensate is
believed to vanish (in the chiral limit) at some critical density
$\rho_c$ (this is more or less supported by models but lattice
calculations are not yet available) corresponding to chiral phase
transition, so one may think of the mass as an indicator for
chiral properties of dense matter. This is currently a hot topic
with considerable controversy.
\subsection{Nuclear matter}
\indent\indent
 The first question one must answer for the viability of the theory
 is: Can this theory describe
 nuclear matter correctly? If the answer were no, then the theory should
 be abandoned. The answer comes out to be yes: the
theory that gives the correct properties of nuclear matter is a
Walecka mean-field-type theory with the nucleon $N$,
isoscalar-vector $\omega_\mu$, and scalar $\sigma$ fields that are
coupled linearly. The Lagrangian is of the form of a linear Walecka
model~\cite{walecka} with, however, the masses of the particles
scaling a la BR (\ref{BR}). It has been thought for some time that
such a theory with the BR scaling would not give a stable system,
not to mention a correct binding energy, saturation density
$\rho_0$ and compression modulus $K$. But this (thinking) turns out
not to be correct. In a Seoul National University
PhD thesis work that succeeds to resolve
some of the long-standing problems, Chaejun Song (presently a post-doc
at SUNY, Stony Brook) -- with help from his senior
collaborators~\cite{song} -- has shown that to correctly interpret
the theory, it is essential to express the density dependence of
the masses in terms of certain chiral-invariant fermion bilinears.
The so-called ``rearrangement terms" do come out correctly for the
resulting equation of state. All properties of nuclear matter are
found to be satisfactorily reproduced while maintaining consistency
with all thermodynamic properties. This gives the assurance that
one should be able to make small fluctuations around the ground
state using the effective Lagrangian, the equilibrium minimum
representing the fixed point.
\subsubsection{Evidence in heavy nuclei}
\indent\indent
 There are some experimental data already available in the literature
 that we can use to test certain  aspect of the BR scaling in
 fluctuations around the equilibrium density. A number of cases
 are available but I shall pick two here for illustration.
\begin{itemize}
\item
 The first is the ``anomalous" orbital gyromagnetic ratio $\delta g_l$
 in heavy nuclei. The orbital gyromagnetic ratio $g_l$ is the
 coefficient figuring in the convection current for a nucleon
 sitting on the Fermi surface responding to slowly varying
 electromagnetic field:
 \bea
 \vec{J}=g_l (e\vec{p}/m_N).\label{J}
 \eea
 Because of the many-body interactions, $g_l$ has an anomalous term $\delta
 g_l$,
 \bea
 g_l=\frac{1+\tau_3}{2}+\delta g_l.\label{gl}
 \eea
 Note that the current (\ref{J}) carries the ``bare"
 mass $m_N$, not the effective quasiparticle (or Landau) mass
 $m^{eff}_N$. Thus the first term of (\ref{gl}) correctly
 describes charge conservation. This is the analog to Kohn's
 theorem in electronic systems~\cite{kohn} and makes the
 calculation of $\delta g_l$ a highly constrained one providing
a stringent consistency condition. Indeed the ``mapping" of
 the BR scaling Lagrangian theory to Landau Fermi-liquid fixed
 point theory gives a highly non-trivial result as shown in~\cite{FR}:
 \bea
 \delta g_l=\frac 49 [\Phi^{-1}-1-\frac 12 \tilde{F}_1^\pi]\tau_3
 \eea
 where  $\Phi$ is as given in (\ref{BR}) and $\tilde{F}_1^\prime$
 is the pion contribution to the Landau $F_1$ parameter which is
 completely determined by chiral Lagrangian. At the nuclear matter
 density $\rho=\rho_0$, both $\Phi$ and $\tilde{F}_1^\prime$ are known
 numerically,
 \bea
 \Phi (\rho_0)= 0.78, \ \ \tilde{F}_1^\prime (\rho_0)=-0.459.
 \eea
 Since heavy nuclei must have $\rho\sim \rho_0$,
 the prediction in heavy nuclei is
\bea
 \delta g_l \approx 0.23\tau_3.
\eea
 This prediction is consistent with the experimental data $\delta
 g_l^{proton}=0.23\pm 0.03$ extracted from giant resonances in the
 lead region. It is also consistent with magnetic moments in the
 lead region.
\item
 Another case that provides support for the scheme is the axial charge
 transitions in heavy nuclei $A(0^\pm)\rightarrow A'(0^\mp) + e^+
 (e^-) +\nu (\bar{\nu})$ with change of one unit of isospin. As shown
 in the PhD thesis (at Seoul National University) of Tae-Sun
 Park~\cite{TSP}, this  particular transition is highly enhanced
 in nuclei (by a factor of $\sim$ 2 with respect to the
 single-particle strength)
 due to one soft pion being exchanged between two nucleons that are
 involved in the response to the axial charge operator. In heavy
 nuclei, this is further enhanced because of the scaling
 $f_\pi/f_\pi^* =\Phi^{-1}>1$ at nuclear matter density~\cite{KR}
 which can be seen as one goes up in mass number.
 Experiments are available in medium and heavy nuclei where this
 enhancement has been seen and confirmed
 unambiguously~\cite{warburton,rho}.
 \end{itemize}

\subsection{Heavy-ion collisions and in-medium meson properties}
 \indent\indent
 Up to here, I have indicated how the BR scaling works for the
 nucleon mass and, indirectly, for the pion decay constant.
 So far the
 scaling for the meson masses has not been tested although it
 figures indirectly in both cases discussed above.
 Heavy-ion collisions could provide a qualitative test of the
 behavior of mesons in dense medium.

 Heavy-ion
 collisions produce hot and dense matter. The physics of the
 process must therefore be able to probe the behavior of the relevant
 degrees of freedom which are mainly mesonic at high temperature
 and/or high density. This is a big area and much debate has been
 going on. A recent review can be found in \cite{rappwambach}.
In a recent 
PhD thesis work, Youngman Kim of Hanyang University (who is currently
a post-doc at SUNY, Stony Brook and University of South Carolina, 
Columbia) has shown that both scalar and vector mesons in hot and dense medium
do indeed scale a la BR scaling
if trace anomaly and hidden gauge symmetry of QCD are properly taken into
account~\cite{YK}.
\begin{figure}[t]
\begin{center}
\epsfig{file=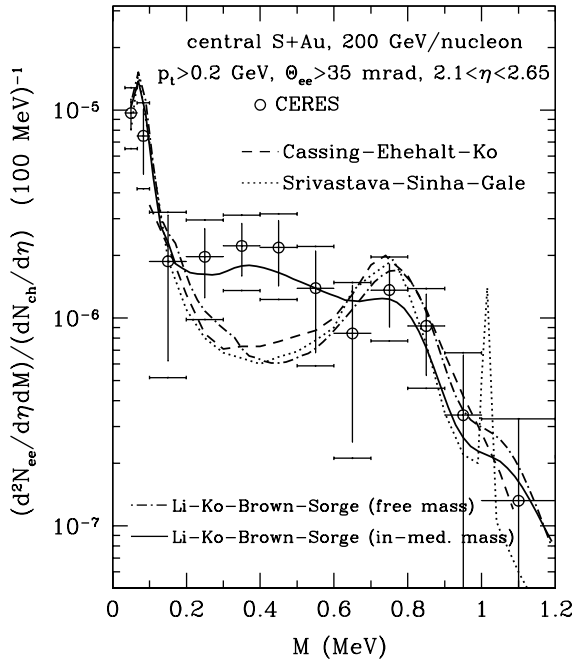,height=4in} \caption{CERES data~\cite{ceres}
and fits~\cite{LLB}}\label{ceres}
\end{center}
\end{figure}

 Without going into the details of the dynamics involved -- which goes
 out of scope for this note, it is
 difficult to be precise about what we are dealing with. So let me
 be glib about it and just show the result and
 give some (perhaps biased) remarks.

 In fig.(\ref{ceres}) is given the CERN (CERES) experiment for $200$ GeV
 per nucleon central collision of $S$ on $Au$ producing dileptons
 measured at CERN. The differential cross section is plotted vs.
 invariant mass $M$. Now
 if one takes the masses of the particles involved in the
 collision to be those of free-space particles, workers in the
 field more or less agree on the predicted cross sections. Despite
 the large error bars for the experiments, one can see that the
 free-mass description largely under-estimates the cross section for
 around $M\sim 400$ MeV.  As shown by Li, Ko and Brown~\cite{LKB},
 the data can be explained quite economically with the
 BR-scaling masses, the primary agent for this being the dropping
 mass of the $\rho$ meson which plays the principal role in the
 dilepton process. A similar fit is obtained in the
 $Pb$-on-$Au$ process.

Unfortunately, this simple picture is blurred by a controversy on
the precise cause for the shift of the peak in the dilepton data.
It appears at present that this BR-scaling mechanism is not the
only one that could explain the data. There are various other
(alternative?) explanations such as the increased width of the
vector meson~\cite{rappwambach} -- thereby possibly invalidating
the quasiparticle interpretation -- or nonperturbative quark-gluon
plasma effect~\cite{leezahed} etc. Whether or not all these
alternatives represent different physical phenomena is not known.
In any event, it would be premature to conclude that the
quasiparticle description a la BR is invalidated. I would say that
whether or not a quasiparticle picture for hadrons is applicable in
dense medium is an open question that cannot be settled by a
few-order calculation in a strong-coupling situation. In fact in
condensed matter physics, there are cases where low-order
treatments fail to give the correct Fermi liquid structure, the
latter resulting only when all-order calculations are performed.
For an example, see \cite{luttinger}.
\subsection{Running kaon mass}
\indent\indent
 Let me now turn to fluctuations around the ground state of
 nuclear matter in other flavor directions than the up- and
 down-quark flavors, say, in the strangeness flavor direction.
 For this
 we need to extend the flavor space to $SU(3)$. We know how to
 do this.

 \begin{figure}[ht]
 \begin{minipage}[b]{0.46\linewidth}
\begin{center}
\epsfig{file=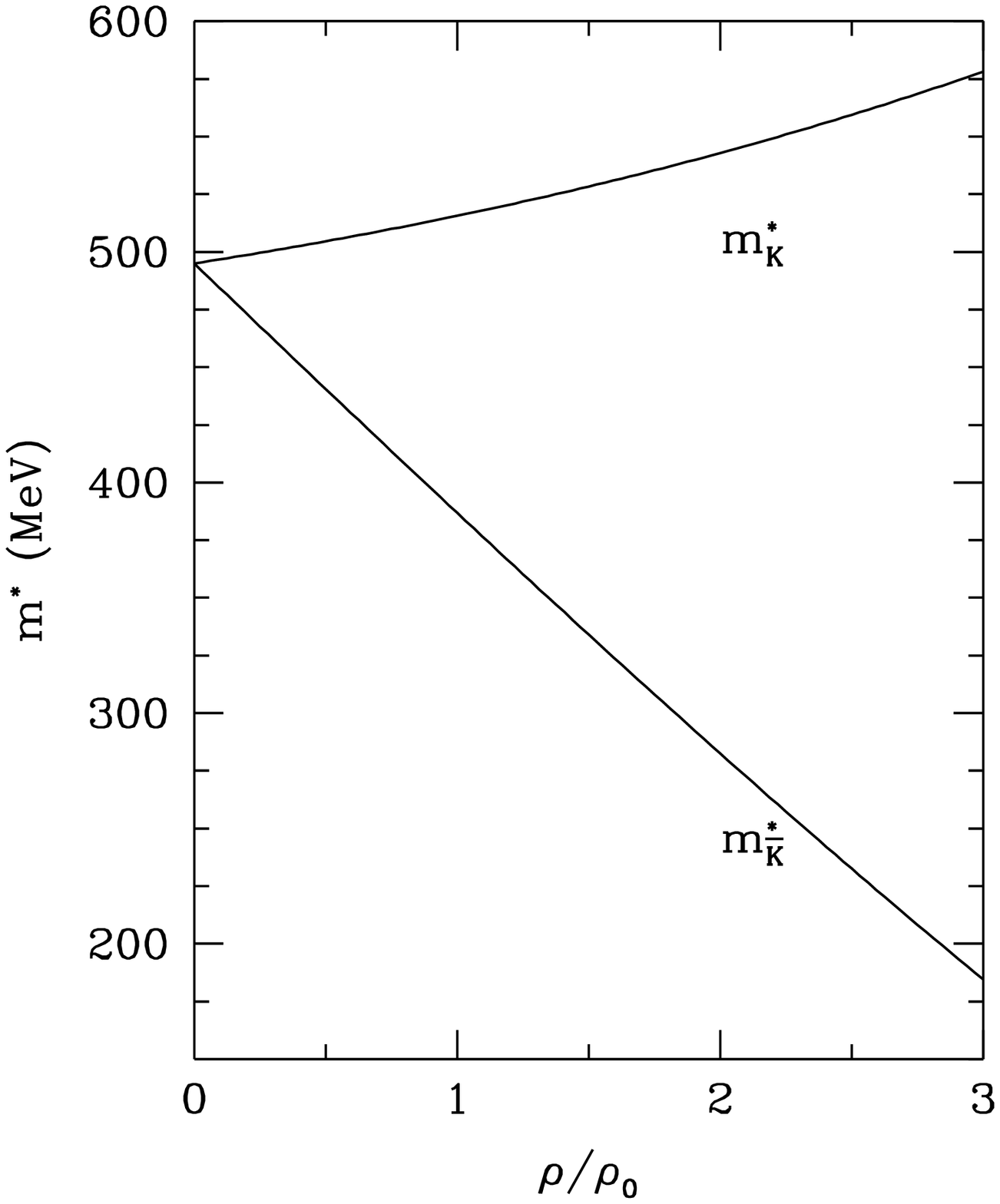,height=3.2in} \caption{In-medium kaon
($K^-$) and anti-kaon ($K^+$) masses calculated with the
empirically constrained dispersion formula~\cite{LLB}.
}\label{kaonmass}
\end{center}
\end{minipage}\hfill
\begin{minipage}[b]{0.46\linewidth}
\begin{center}
\epsfig{file=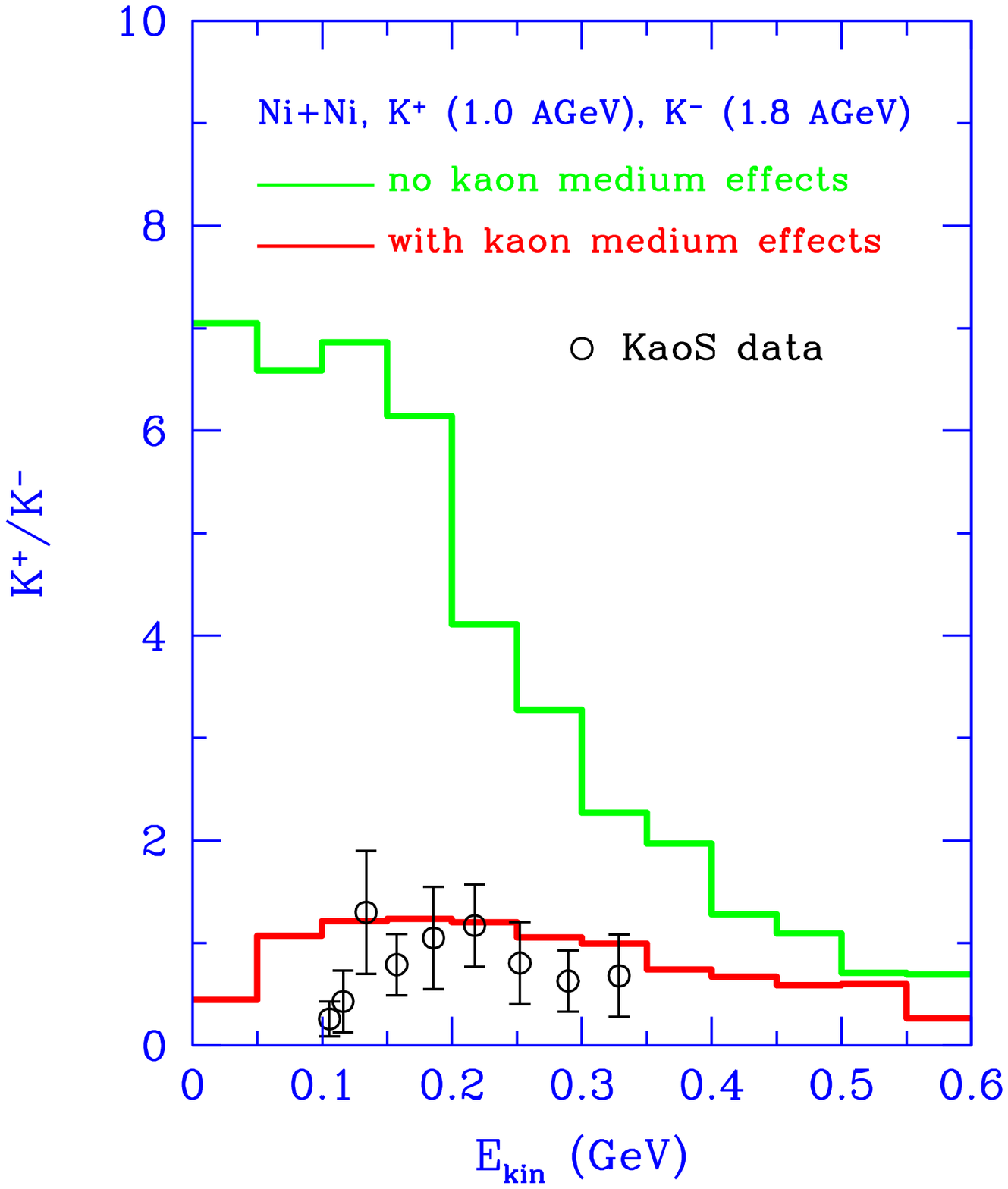, width=3.2in}\caption{Kinetic energy spectra
of $K^+/K^-$ in Ni + Ni collisions. The circles are experimental
data from the KaoS collaboration~\cite{kaos}.}\label{exp-th}
\end{center}
\end{minipage}
\end{figure}

 A kaon propagating in dense medium interacts with the background
 with its  mass and coupling modified by the medium. The coupling
 with the matter field (nucleons) is given
 by $\sim {f^*}^{-1}$ where $f^*$ is the in-medium
 pseudo-Goldstone boson  (pion to the leading order) decay
 constant which scales with density. By (\ref{BR}), the scaling
 coupling can be related to the scaling in the masses of the
 mesons that are exchanged between kaon and nucleon in a
 description where heavy mesons are explicitly accounted for.

 The masses for $K^{\pm}$ so predicted are plotted in
 fig.(\ref{kaonmass}). Actually what is calculated here is
 the kaon dispersion relation with inputs from experiments
 which is equivalent, to leading order, to using a BR-scaled
 chiral Lagrangian in the tree order. This and related matters
 are discussed in the PhD thesis  work~\cite{leePR} of
 Chang-Hwan Lee at Seoul National University (presently at SUNY,
 Stony Brook). This prediction has been beautifully checked by
 experiments~\cite{LLB} as shown in fig.(\ref{exp-th}).

\subsection{Kaon condensation and neutron stars}
\begin{figure}
 \begin{minipage}[b] {0.46\linewidth}
 \begin{center}
 \epsfig{file=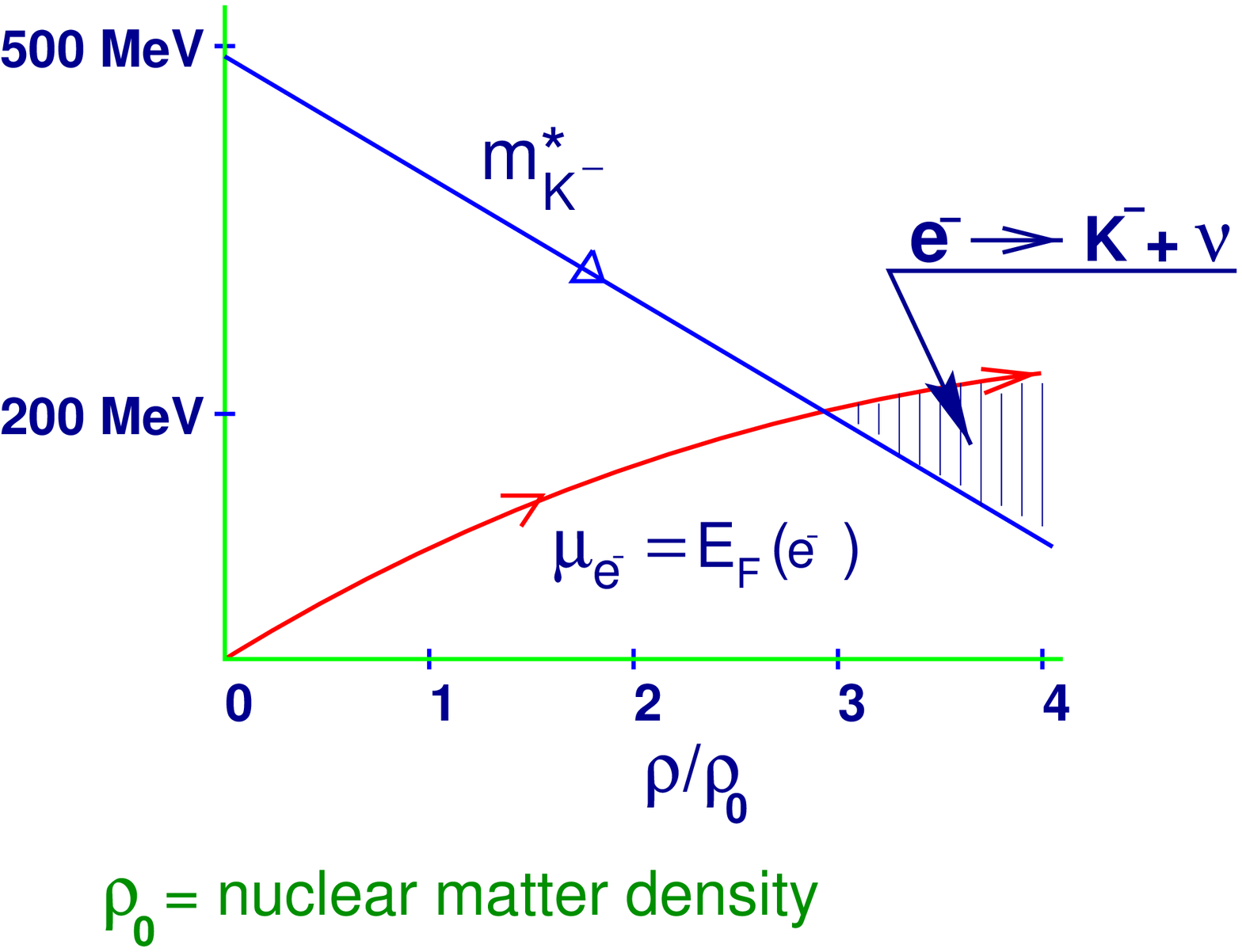, width=3.2in} \caption{Electron
 conversion to $K^-$ and kaon condensation.}\label{kcon}
 \end{center}
\end{minipage}\hfill
\begin{minipage}[b]{0.46\linewidth}
\begin{center}
\epsfig{file=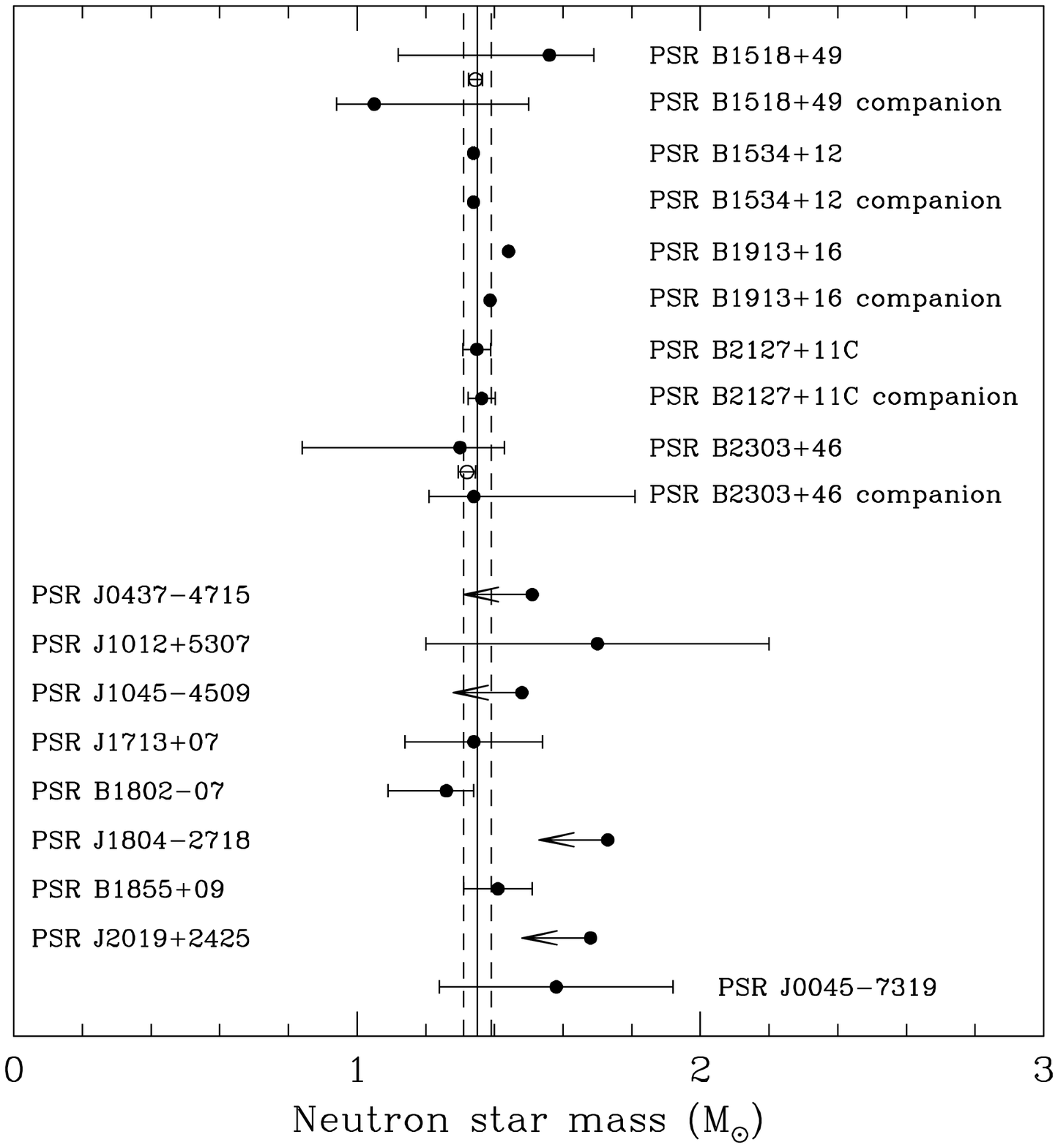,width=3.2in}
 \caption{Measured neutron star masses~\cite{nmass}.}\label{nstarmass}
 \end{center}
 \end{minipage}
  \end{figure}

\indent\indent
 An important consequence of the dropping $K^-$ mass is its
 effect on the equation of state for dense neutron matter and
 consequently on neutron stars. This is the link between hadron
 physics and astrophysics that has been emphasized by Gerald E. Brown
 and Chang-Hwan Lee. The basic idea is quite simple. In
 dense medium, the $K^-$ mass drops continuously as density
 increases. But as density increases, the electron chemical
 potential $\mu_e$ in neutron-star matter increases. When the
 effective kaon
 mass crosses the electron chemical potential, the electrons can
 turn into kaons which by nature of their bosonic character can
 condense as first suggested by Kaplan and Nelson~\cite{nelson}.
 This is shown schematically in fig.(\ref{kcon}).

 The onset of kaon condensation is expected to soften the equation of
 state of
 the matter. The intricate interplay between this phenomenon,
 hadronic interactions and gravitational  interactions has been worked
 out by Li, Lee and Brown~\cite{LLB}.
Their finding is that the maximum mass of neutron stars can be
lowered by about 0.4$M_\odot$, once kaon condensation as
constrained by the dropping kaon mass -- reflected in the
empirical dispersion relation -- is introduced. The authors point
to ``the growing interplay between hadron physics, relativistic
heavy-ion physics and the physics of compact objects in
astrophysics." This may provide a natural explanation (hopefully
without fine-tuning) for the observation that the well-measured
neutron-star masses fall within a narrow window $M\lsim 1.6
M_\odot$ (see fig.\ref{nstarmass}~\cite{nmass}).)

\section{Superdense Matter and The Cheshire Cat}
\indent\indent
 The last topic I would like to discuss is infinite
nuclear matter at large density, a density much greater that
considered above. This may be relevant to neutron star cooling
although at present it is not clear whether at the density
appropriate for neutron star matter other processes cannot compete
with it~\cite{kurt,PRWZ}.

 The old lore that at an asymptotic density, the
matter can be described by perturbative QCD with weakly
interacting boring quarks is now widely recognized to be simply
wrong. What may be happening at super-high density is something a
lot more intriguing and exciting than previously thought. This
explains a flurry of activity throughout the world, including the
present intense activity at Korea Institute for Advanced Study
KIAS) in which I have been participating.

What is most surprising and in some sense unexpected is that at
high density the Cheshire Cat picture re-emerges! In fact, at high
density, there ceases to be any real distinction between quarks and
hadrons. This can be best seen in terms of a quark soliton
analogous to the qualiton Kaplan~\cite{kaplan} introduced as a
model for the constituent quark. The mechanism I will discuss
exploits that at high density diquarks condense giving rise to
color superconductivity as proposed some years ago~\cite{bailin} and
recently revived \cite{CSC}. Since the resulting qualiton is formed
from a color superconducting ground state, it seems proper to call
it superqualiton~\cite{superqualiton}. It has been argued on
general symmetry and dynamical grounds~\cite{duality} that at high
density, hadronic matter of flavor $SU(3)$ is characterized by the
condensate
\be
\Big< q_{L\alpha}^{ia}q_{L\beta}^{jb} \Big>
=-\Big<q_{R\alpha}^{ia}q_{R\beta}^{jb}\Big> =\kappa
\,\,\epsilon^{ij}\epsilon^{abI}\epsilon_{\alpha\beta I} \label{1}
\ee
 where $\kappa$ is some constant, $i,j$ are $SL(2,C)$ indices,
$a,b$ are color indices, and $\alpha,\beta$ are flavor indices.
Equation (\ref{1}) holds for parity-even states. Such a condensate
locks color and flavor so that global color and chiral symmetry
are broken to the diagonal subgroup $SU(3)_{C+L+R}$~\footnote{I am
describing here the situation where the color and flavor lock in
such a way that the color symmetry is completely broken. Other
cases where color is partially broken with differing flavor
patterns could also be addressed in a similar way using a qualiton
picture although details remain to be worked out.}. The
consequence of this is that there is an invariant $U(1)$ subgroup
that contains a ``twisted" photon, measured with which all
excitations carry integer charges reminiscent of the Han-Nambu
quarks and have quantum numbers that correspond to those of the
mesons and baryons present at zero density. There is then a
continuity between the excitations at high density in terms of
quarks and gluons and hadronic excitations at low density in terms
of baryons and mesons. This clearly is a case of Cheshire Cat.

Now to see that this is the Cheshire Cat in the sense formulated in
terms of the chiral bag~\cite{NRZ}, consider the excitation of a
quark on top of the diquark-condensed ``vacuum." In \cite{duality},
such a quark is argued to behave like a baryon. Now I claim that
this quark is a quark soliton, i.e.,
superqualiton~\cite{superqualiton}.

To describe the low-energy dynamics of the color-flavor locking
phase, introduce a field $U_L(x)$ which maps space-time to the
coset space, $M_L=SU(3)_c\times SU(3)_L/SU(3)_{c+L}$. One can take
it to be
\begin{equation}
{U_L}_{a\alpha}(x)\equiv\lim_{y\to x}{\left|x-y\right|^{\gamma_m}
\over\kappa}\epsilon^{ij}\epsilon_{abc}\epsilon_{\alpha\beta\gamma}
q^{b\beta}_{Li}(-\vec v_F,x)q^{c\gamma}_{Lj}(\vec v_F,y),\label{U}
\end{equation}
where $\gamma_m$ is the anomalous dimension of the diquark field
of order $\alpha_s$ and $q(\vec v_F,x)$ denotes the quark field
with momentum close to a Fermi momentum $\mu\vec v_F$. The pairing
involves quarks near the opposite edge of the Fermi surface.
Similarly, we introduce a right-handed field $U_R(x)$, also a map
from space-time to $M_R=SU(3)_c\times SU(3)_R/SU(3)_{c+R}$, to
describe the excitations of the right-handed diquark condensate.
If this field takes a vacuum expectation value as a consequence of
the diquark condensation which will, owing to (\ref{1}), have the
form \bea
\left<{U_L}_{a\alpha}\right>=-\left<{U_R}_{a\alpha}\right>=\kappa\,
\delta_{a\alpha}, \eea then 16 Nambu-Goldstone bosons will get
excited~\footnote{Actually there are 17 of them, one of which
having to do with spontaneous breaking of the baryon number.}.
Eight of them will get eaten up by the gluons to give masses to
the gluons. The massive gluons then turn into massive vector
mesons whose quantum numbers are those of the light-quark vector
mesons present at zero density. The remaining eight (pseudoscalar)
Nambu-Goldstone bosons are the equivalents of the ones present at
zero density and are represented by the interpolating field
(\ref{U}). In analogy to the usual skyrmion at zero density, this
field supports a soliton which is a fermion, the quantum numbers
of which are identical to those of the usual baryon.

The effective Lagrangian that gives rise to this soliton should in
principle be derived from QCD. At the moment such an effective
Lagrangian is not known. However one can venture to make a few
interesting conjectures. Viewed as a superqualiton whose mass is
given by the soliton mass, there is nothing that requires that the
soliton mass be equal to or near the superconductivity gap
$\Delta$ (which is dictated by the condensate). In fact there is
nothing which would prevent the mass from being much less than the
gap. Thus one could imagine that light fermions are excited {\it
within} the gap. Correlations between light superqualitons could
rearrange the ground state into a different form from that of the
standard superconductivity. There could be other modes of similar
nature such as particle-hole excitations from the opposite ends of
the Fermi sea (much like the Cooper pair but involving particles
and holes) which could give rise to a crystalline structure
etc.~\cite{PRWZ}. For this and other reasons the phenomenon of
color superconductivity in QCD at high density could be completely
different from the usual BCS superconductivity. The issue of how
this matter could influence the structure of compact stars (e.g.,
cooling, equation of state etc.) is an open one actively studied
presently in KIAS~\cite{kias}.

\section{Conclusion} \indent\indent The most important outcome of
the recent development of EFT in nuclear physics is that the
highly successful approach to nuclear structure using realistic
nuclear potentials (PM) is rendered a first-principle
interpretation in that it represents the leading term in the EFT
expansion with the corrections thereof systematically calculable.
This confers the power of modern field theory techniques to the
standard nuclear physics approach that has been practiced with
success since a long time. This ``bridging" comes about thanks to
a possible duality that I refer to as Cheshire Cat Principle
between QCD variables and macroscopic (color-singlet) variables.
This also provides a potential link between the physics of the
elementary nucleon, nuclei, hadronic matter and compact-star
matter. In the case of high density, the picture becomes even more
intriguing. There we see emerging the symbolic (approximate)
equality
 \bea
  ``{\rm Quark}"\approx ``{\rm
Qualiton}"\approx ``{\rm Baryon}".
\eea

 It is amusing that the notion of the Cheshire Cat which
was conceived by the need to reconcile the traditional
meson-exchange description with the modern QCD description for
nuclear processes~\cite{littlebag} (i.e., the ``little bag" with
pion cloud, chiral bag etc) at low density re-emerges at high
density where one would have expected the bona-fide QCD to be
uniquely applicable.

Most significant of all, the interplay between hadronic physics,
relativistic heavy-ion physics and the physics of compact stars in
astrophysics highlights the unity of physics, an endeavor that
could make a mainstream of physics research in Asia Pacific
research centers like APCTP and in Korean institutes like KIAS.
Such a potential seems only natural given that some of the most
significant contributions have been, and are being, made by young
Korean theorists -- graduate students and post-docs -- actively
working in the field and not less significantly
that several experimental collaborations 
between Korean experimenters and the ALICE (CERN) and RHIC (Brookhaven)
project teams purporting to probe the hot and dense matter relevant to
early Universe and compact stars are in the process of being formed.

\subsection*{Acknowledgments}
\indent\indent I am very grateful to Chang-Hwan Lee for his
invaluable help in preparing this note and for useful
comments. I would like to thank also
all those who participated this year in the working group on
various aspects of dense hadronic matter, QCD and astrophysical
consequences at KIAS, in particular, Deog Ki Hong, Steve Hsu, 
Kuniharu Kubodera, Kurt
Langfeld, Hyun Kyu Lee, Dong-Pil Min, Byung-Yoon Park, Maciek Nowak, 
Sang-Jin Sin, Vicente Vento and Ismail Zahed.

\end{document}